\documentclass[conference,a4paper]{APSIPA2026}
\IEEEoverridecommandlockouts
\usepackage{amsmath,amssymb}
\usepackage{graphicx}
\usepackage{multirow}
\usepackage{threeparttable}
\usepackage[backend=biber,style=ieee,]{biblatex}
\usepackage{geometry}
\usepackage{fancyhdr}

\fancypagestyle{firststyle}{
  \fancyhf{}
  \fancyhead[C]{2026 Asia Pacific Signal and Information Processing Association Annual Summit and Conference (APSIPA ASC)}
}

\usepackage{adjustbox}
\usepackage{makecell}
\usepackage[hidelinks]{hyperref}

\begin{document}

\title{Deep Neural Compression for RIR-Characterized Acoustic Environments with Structure-Aware Constraints}

\author{
\authorblockN{
Chen-Yuan Ning,
Yang~Ai\authorrefmark{1},
Hui-Peng Du,
Xiao-Hang Jiang,
and Zhen-Hua Ling
}
\authorblockA{
National Engineering Research Center of Speech and Language Information Processing,\\
University of Science and Technology of China, Hefei, China \\
E-mail: \{chenyuanning,redmist,jiang\_xiaohang\}@mail.ustc.edu.cn, \{yangai,zhling\}@ustc.edu.cn
}

\thanks{\authorrefmark{1}\ Corresponding author. This work was funded by the National Natural Science Foundation of China under Grant 62301521.}
}
\maketitle
\thispagestyle{firststyle}
\pagestyle{empty}

\begin{abstract}
Room impulse responses (RIRs) characterize the acoustic environment of a room by capturing how sound propagates and decays within an enclosed space. 
In applications such as immersive audio rendering, accurate acoustic reconstruction often relies on spatially densely sampled RIRs. 
This consequently gives rise to a large volume of RIR data, imposing a substantial burden on storage.
Although recent neural audio codecs provide an effective framework for low-bitrate compression, their training objectives are mainly tailored to speech and general audio, and are therefore not well aligned with the acoustic characteristics of RIRs.
Therefore, we propose an EnCodec-based neural RIR compression method, which incorporates RIR structure-aware constraints at two levels. 
Specifically, at the RIR level, structure-aware constraints are imposed on the global decay behavior and local energy distribution of RIRs through energy decay curve (EDC) regularization and a short-time window energy constraint, while at the reverberant-speech level, reverberant-speech supervision is further introduced to constrain the consistency of the reverberant speech generated by the reconstructed RIRs. 
Experimental results show that, at a low bitrate of 375~bps, the proposed method achieves lower RIR reconstruction error and better reverberant-speech perceptual consistency than audio-oriented codecs.
\end{abstract}

\section{Introduction}
\label{sec:intro}

Room impulse responses (RIRs) characterize the acoustic environment of a room under a specific source–receiver configuration by capturing how sound propagates in an enclosed space and interacts with the boundaries~\cite{schissler2017srir}.
RIRs from the same acoustic environment often need to be repeatedly accessed and reused in applications such as immersive audio rendering~\cite{lan2024acoustic,geldert2023srir}, room acoustic analysis~\cite{iso2009acoustics}, and speech enhancement~\cite{benesty2006speech,gannot2017multimicrophone}. 
Instead of repeatedly measuring, re-modeling, or reconstructing the same acoustic environment, it is often more practical and efficient to directly store measured RIRs for future reuse. 
However, as the number of spatial sampling points increases, the number of RIRs that need to be stored also grows rapidly. 
Moreover, since each RIR is typically sampled at a high rate and often spans a nontrivial temporal duration, large-scale RIR datasets can impose substantial burdens on storage and management.
Efficient RIR compression is therefore of clear practical importance for real-world deployment.

Prior studies on RIR compression have explored low-rank-based representations, including joint compression using generalized low-rank approximation of matrices (GLRAM)~\cite{jalmby2024multi}, as well as simpler approximation strategies such as truncation and thresholding~\cite{jalmby2024compression}.
While these approaches have shown clear benefits for compact storage and low-latency processing, jointly achieving high compression efficiency and faithful preservation of acoustically important characteristics remains challenging.

In the field of audio processing, codecs play a fundamental role in reducing the storage cost of digital audio by compressing raw waveforms into compact discrete representations with minimal reconstruction distortion.
In recent years, neural audio codecs built upon end-to-end encoder-quantizer-decoder architectures have demonstrated strong compression performance, showing a remarkable ability to preserve perceptual quality at low bitrates~\cite{zeghidour2021soundstream,defossez2022high,yang2023hifi,kumar2023high,ai2024apcodec,jiang2024mdctcodec,liu2024semanticodec}.
However, they were primarily developed for speech or general audio, with training objectives mainly emphasizing waveform or spectral reconstruction fidelity and perceptual quality. 
At present, some researchers have begun to explore the direct use of off-the-shelf neural audio codecs for RIR compression, with results showing that the official pretrained EnCodec \cite{defossez2022high} can achieve low-bitrate RIR compression to some extent~\cite{mezza2024largescale}. 
However, the overall performance remains limited, suggesting that directly applying existing audio codecs to RIR compression is suboptimal. 
A likely reason is that their design objectives are primarily tailored to speech and general audio, rather than the distinctive usage patterns and structural characteristics of RIRs.


In view of the distinctive temporal structure of RIRs, we propose a neural RIR compression method based on the EnCodec backbone, with joint constraints introduced at both the RIR level and the reverberant-speech level.
Regarding the RIR-level constraints, besides the conventional audio-oriented compression objectives, an energy decay curve (EDC) constraint over the effective decay interval and a short-time window energy constraint are introduced to preserve the global decay behavior and local temporal energy distribution of RIRs. 
Regarding the reverberant-speech-level constraints, reverberant-speech supervision is introduced to constrain the discrepancy between the reverberant speech signals generated from the reference and decoded RIRs in both the time domain and mel domain, thereby improving the consistency of downstream reverberant speech and indirectly facilitating RIR learning. 
Experimental results on the public real-world RIR dataset Motus~\cite{gotz2021dataset} at 24~kHz confirm that the proposed method can enable efficient and high-quality RIR compression. 
At a low bitrate of only 375~bps, it achieves the lowest T60 error for RIR reconstruction and the highest virtual speech quality objective listener (ViSQOL) score~\cite{chinen2020visqol} of 4.11 for the generated reverberant speech among the compared audio-oriented codec baselines.

\section{Proposed Method}
\label{sec:method}

\begin{figure}[t]
    \centering
    \includegraphics[width=1\linewidth]{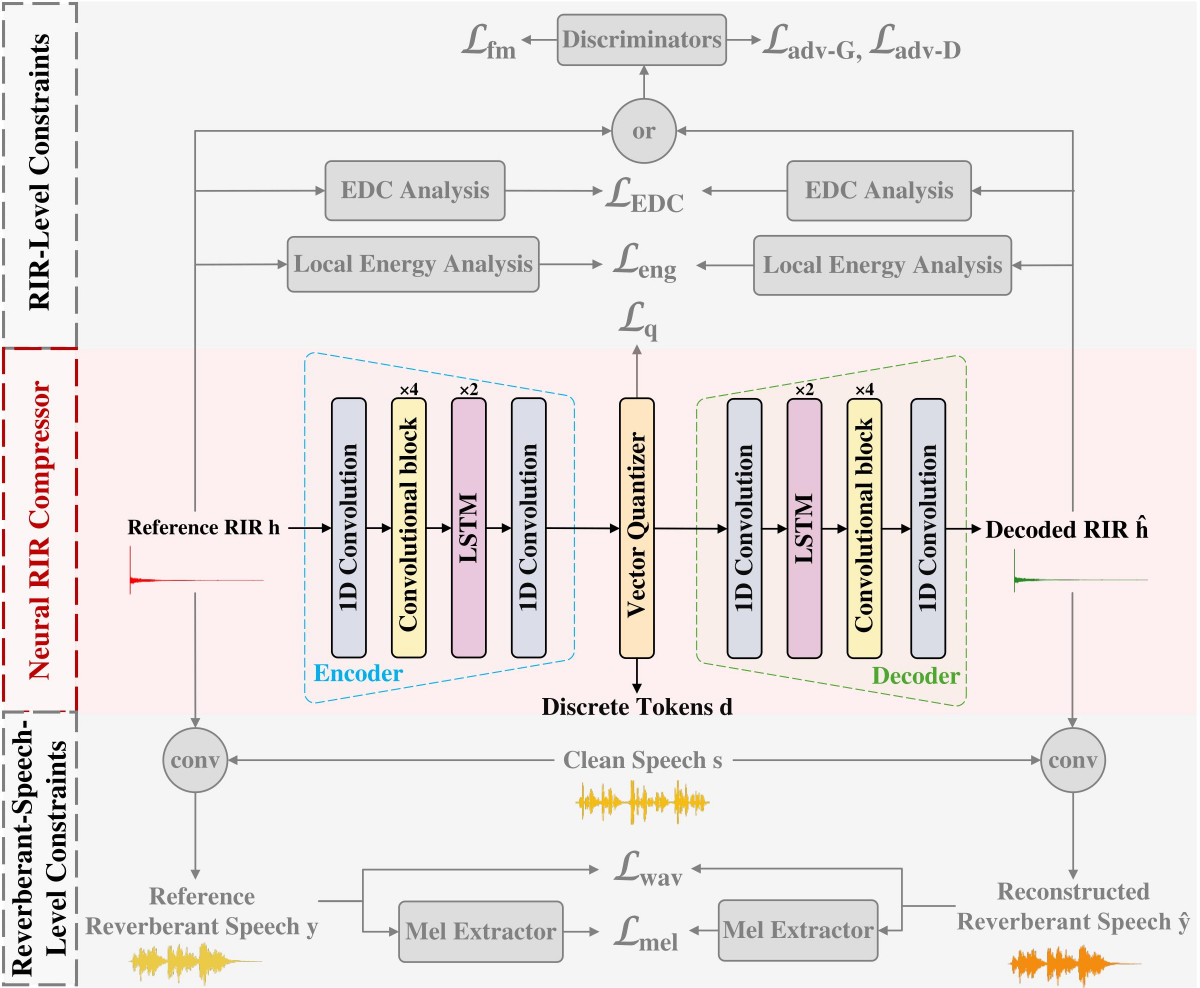}
    \caption{An overview of the proposed neural RIR compression method.}
    \label{fig:1}
\end{figure}

An overview of the proposed neural RIR compression method is shown in Fig.~\ref{fig:1}. 
This method compresses an input RIR sequence $\mathbf{h} = [h(0), h(1), \ldots, h(N-1)]^{\top} \in \mathbb{R}^{N}$ into a discrete token sequence $\mathbf{d} = [d(0), d(1), \ldots, d(L-1)]^{\top} \in \mathbb{N}^{L}$ for compact storage, and the RIR sequence can then be reconstructed from the discrete tokens as $\hat{\mathbf{h}} = [\hat{h}(0), \hat{h}(1), \ldots, \hat{h}(N-1)]^{\top} \in \mathbb{R}^{N}$ through the decoder, where $N$ and $L$ are the lengths of the RIR sequence and the discrete token sequence. 
For model training, the proposed method adopts RIR-structure-aware constraints at both the RIR level and the reverberant-speech level. 
Details of the model architecture and training strategy of the proposed method are described as follows.


\subsection{Model Details of Neural RIR Compressor}
\label{sec:II-A}

As shown in Fig.~\ref{fig:1}, the proposed neural RIR compressor follows the general design of EnCodec \cite{defossez2022high}, consisting of a convolution-based encoder, a quantizer, and a convolution-based decoder. 
Taking the RIR sequence as input, the encoder starts with a one-dimensional convolution layer with 32 channels and a kernel size of 7, followed by 4 convolutional blocks.
Each block first applies a residual unit, which consists of two one-dimensional convolution layers with kernel sizes of 3 and 1, respectively, together with a skip connection. 
In the residual unit, the channel dimension is first reduced to half of the input channels and then restored to the original size. 
This is followed by a downsampling layer implemented as a strided one-dimensional convolution whose kernel size is set to twice the corresponding stride, and whose output channel dimension is doubled relative to the input.
The strides of the 4 downsampling layers are set to 4, 4, 5, and 8, respectively, which together determine the overall downsampling ratio of the encoder. 
After the 4 convolutional blocks, a two-layer long short-term memory (LSTM) network, with 512 hidden units in each layer, is used for sequence modeling, followed by a final one-dimensional convolution layer with a kernel size of 7 and 512 output channels.
The decoder is symmetric to the encoder, using transposed convolutions in place of strided convolutions and adopting the reverse stride order of the encoder.


Unlike EnCodec, which adopts residual vector quantization, our neural RIR compressor employs a single-codebook vector quantizer (VQ) to achieve deep, low-bitrate compression. 
Assuming that the codebook size is $M$, the bitrate of the discrete tokens can be calculated as $\frac{s_r}{R}\cdot \log_2 M$ bps, where $s_r$ is the sampling rate and $R$ denotes the waveform-to-token downsampling ratio, which is equal to the product of the strides of all downsampling layers.


\subsection{Structure-Aware Multi-Level Constraints}

The training of the neural RIR compressor adopts RIR-structure-aware constraints at both the RIR level and the reverberant-speech level. 

\subsubsection{RIR-Level Constraints}

The RIR-level constraints are designed to reduce the discrepancy between the decoded RIR sequence $\hat{\mathbf{h}}$ and the reference RIR sequence $\mathbf{h}$ from multiple aspects according to the structural characteristics of RIRs, including the following losses.

\textbf{Energy Decay Curve Loss.}
The EDC characterizes the global decay of acoustic energy over time and can be derived from an RIR via Schroeder's backward integration~\cite{schroeder1965new}. 
To preserve the reverberation decay characteristics of reconstructed RIRs, we introduce an EDC-based structure-aware loss to quantify the discrepancy between the reference and reconstructed RIRs in terms of their global decay behavior.
In contrast, the modeling and reconstruction of general audio signals typically do not explicitly emphasize such global decay characteristics, and thus rarely impose a dedicated constraint on the EDC error. 
For RIR $\mathbf{h}$ and $\mathbf{\hat{h}}$, the backward-integrated energy is defined as
\begin{equation}
\varepsilon_h(n)=\sum_{k=n}^{N-1} h^2(k),\quad \varepsilon_{\hat{h}}(n)=\sum_{k=n}^{N-1} \hat{h}^2(k),
\end{equation}
where $n=0,1,\dots,N-1$. 
The corresponding normalized logarithmic EDC is given by
\begin{equation}
e_h(n)=10\log_{10}\frac{\varepsilon_h(n)+\epsilon}{\varepsilon_h(0)+\epsilon}, \quad e_{\hat{h}}(n)=10\log_{10}\frac{\varepsilon_{\hat{h}}(n)+\epsilon}{\varepsilon_{\hat{h}}(0)+\epsilon},
\end{equation}
where $\epsilon$ is a small constant for numerical stability. 
The EDC loss is computed as the mean squared error (MSE) between the reference and reconstructed log-EDC curves over the effective decay region, i.e.,
\begin{equation}
\mathcal{L}_{\mathrm{EDC}}=\frac{1}{n^*+1}\left\| \mathbf{e}_h^*-\mathbf{e}_{\hat h}^* \right\|_2^2,
\end{equation}
where $\mathbf{e}_h^* = [e_h(0), e_h(1), \ldots, e_h(n^*)]^{\top} \in \mathbb{R}^{n^*+1}$ and $\mathbf{e}_{\hat h}^* = [e_{\hat h}(0), e_{\hat h}(1), \ldots, e_{\hat h}(n^*)]^{\top} \in \mathbb{R}^{n^*+1}$. 
$n^*$ denotes the upper index of the effective decay interval and is defined by the earlier of the -35 dB crossing points of the reference and reconstructed EDCs, i.e., $n^*=\min(n_h^*,n_{\hat h}^*)$, where $n_h^*=\min\{n\mid e_h(n)\le -35\}$ and $n_{\hat h}^*=\min\{n\mid e_{\hat h}(n)\le -35\}$.


\textbf{Local Energy Loss.}
RIRs exhibit distinct temporal regions corresponding to different acoustic components, including the direct sound, early reflections, and the late reverberant tail~\cite{stewart2007statistical,Vairetti2017OBF}. 
Consequently, different local windows exhibit different energy patterns. 
To preserve the local temporal energy distribution of reconstructed RIRs, we introduce a local energy loss as
\begin{equation}
\mathcal{L}_{\mathrm{eng}}
=
\sum_{j=1}^{J}
\left(
\sum_{n\in\mathcal{W}_j} h^2(n)
-
\sum_{n\in\mathcal{W}_j} \hat{h}^2(n)
\right)^2,
\end{equation}
where $\mathcal{W}_j=\{\,n \mid (j-1)N_w \le n < jN_w\,\}$ denotes the index set of the $j$-th non-overlapping short-time window. 
Here, $N_w=s_r \cdot 0.05$ denotes the number of samples in each 50 ms window, and $J$ denotes the total number of windows.

\textbf{Generative Adversarial Loss.}
Following~\cite{yang2023hifi}, we employ three types of discriminators: a multi-scale STFT discriminator (MS-STFTD), a multi-period discriminator (MPD), and a multi-scale discriminator (MSD). 
The discriminators take either $\mathbf{h}$ or $\hat{\mathbf{h}}$ as input and are trained with hinge-loss objectives. 
The generator loss and discriminator loss are denoted by $\mathcal{L}_{\mathrm{adv-G}}$ and $\mathcal{L}_{\mathrm{adv-D}}$, respectively. 
We also introduce a feature-matching loss $\mathcal{L}_{\mathrm{fm}}$, which is defined on the intermediate-layer features of the discriminators and encourages the reconstructed samples to match the real samples in these feature spaces.


\textbf{Quantization loss}.
Following a similar strategy to \cite{defossez2022high}, we introduce a quantization loss $\mathcal{L}_{q}$ between the encoder output and its quantized representation. 
This term is specifically defined as the MSE between the encoder latent representation and its quantized counterpart, so as to encourage the latent representation to remain close to the selected codeword and improve the stability of quantization.

Overall, the generator-side loss of RIR-level constraints is formulated as
\begin{equation}
\mathcal{L}_{\mathrm{rir}}
=
\lambda_{\mathrm{EDC}}\mathcal{L}_{\mathrm{EDC}}
+\lambda_{\mathrm{eng}}\mathcal{L}_{\mathrm{eng}}
+\lambda_{\mathrm{adv}}\mathcal{L}_{\mathrm{adv\mbox{-}G}}
+\lambda_{\mathrm{fm}}\mathcal{L}_{\mathrm{fm}}
+\lambda_{q}\mathcal{L}_{q}, 
\end{equation}
and the discriminator-side loss of RIR-level constraints is $\mathcal{L}_{\mathrm{adv\mbox{-}D}}$, where $\lambda_{\mathrm{EDC}}$, $\lambda_{\mathrm{eng}}$, $\lambda_{\mathrm{adv}}$, $\lambda_{\mathrm{fm}}$ and $\lambda_q$ are hyperparameters.


\subsubsection{Reverberant-Speech-Level Constraints}

In practical applications, RIR signals are typically used by convolving them with source signals to generate reverberant speech, rather than being perceived directly as standalone waveforms. 
Therefore, the fidelity of reconstructed RIRs should ultimately be reflected in the consistency between the generated reverberant speech and the corresponding real reverberant speech. 
Therefore, we introduce a reverberant-speech consistency loss, including constraints in both the time domain and the mel domain. 
Specifically, given a clean speech signal $\mathbf{s}$, we convolve it with the reference and decoded RIRs, $\mathbf{h}$ and $\hat{\mathbf{h}}$, to obtain the corresponding reference and reconstructed reverberant speech signals, denoted by $\mathbf{y}=\mathbf{s} * \mathbf{h}$ and $\hat{\mathbf{y}}=\mathbf{s} * \hat{\mathbf{h}}$, respectively. 
In the time domain, we minimize the MSE between the reference and reconstructed reverberant speech waveforms, and define the corresponding loss term as 
\begin{equation}
\mathcal{L}_{\mathrm{wav}}=
\frac{1}{N}\left\|\mathbf{y}-\hat{\mathbf{y}}\right\|_2^2.
\end{equation}
In the mel domain, we define the following loss, which is a linear combination of the $L_1$ and $L_2$ losses on the multi-scale mel-spectrograms:
\begin{align}
\mathcal{L}_{\mathrm{mel}}
&=
\sum_{i=1}^I
\left(
\|S_i(\mathbf{y})-S_i(\hat{\mathbf{y}})\|_1
+
\alpha_i \|S_i(\mathbf{y})-S_i(\hat{\mathbf{y}})\|_2
\right),
\end{align}
where $S_i(\cdot)$ denotes the Mel-spectrogram computed under the $i$-th configuration, and $\alpha_i=\sqrt{2^i/2}$ is the weighting factor for linearly combining the $L_1$ and $L_2$ terms at each configuration.


Overall, the loss of reverberant-speech-level constraints is formulated as
\begin{equation}
\mathcal{L}_{\mathrm{rev-spe}}
= \lambda_{\mathrm{wav}} \mathcal{L}_{\mathrm{wav}}
+\mathcal{L}_{\mathrm{mel}}.
\end{equation}
where $\lambda_{\mathrm{wav}}$ is a hyperparameter.

\subsubsection{Overall Training Objective}
The overall training objective is composed of the RIR-level constraints and the reverberant-speech-level constraints, and the neural RIR compressor is trained using an adversarial training strategy. 
Specifically, we use the final generator loss 
\begin{equation}
\mathcal{L}=\mathcal{L}_{\mathrm{rir}}+\mathcal{L}_{\mathrm{rev-spe}},
\end{equation}
and the discriminator loss $\mathcal{L}_{\mathrm{adv\mbox{-}D}}$ to optimize the neural RIR compressor and discriminators alternately.


\begin{table*}[t]
    \centering
    \caption{Objective and subjective evaluation results of the proposed method and baselines. MOS results were reported as mean ± 95\% confidence interval.}
    \label{tab:objective_results}
    \small
    \renewcommand{\arraystretch}{1.12}
    \setlength{\tabcolsep}{4.5pt}
    \adjustbox{max width=\textwidth}{
    \begin{tabular}{cccc|cccc}
        \Xhline{1.2pt}
        \multirow{2}{*}{\textbf{Method}}  
        & \multicolumn{3}{c|}{\textbf{RIR-Level Metrics}} 
        & \multicolumn{4}{c}{\textbf{Reverberant-Speech-Level Metrics}} \\
        \cline{2-4} \cline{5-8}
        & \textbf{T60 Error (s)$\downarrow$}
        & \textbf{SFM Error$\downarrow$}
        & \textbf{DRR Error (dB)$\downarrow$}
        & \textbf{LSD$\downarrow$}
        & \textbf{SegSNR (dB)$\uparrow$}
        & \textbf{ViSQOL$\uparrow$}
        & \textbf{MOS$\uparrow$} \\
        \hline
        Speech-trained EnCodec  
        & 5.04 
        & 0.23 
        & 16.99 
        & 0.76 
        & -6.88 
        & 3.44 
        & 3.00 $\pm $0.11 \\
        RIR-trained EnCodec     
        & 2.49 
        & 0.27 
        & 29.11 
        & 1.03 
        & -8.15 
        & 3.23 
        & 2.32 $\pm $0.12 \\
        Proposed 
        & \textbf{1.14} 
        & \textbf{0.06} 
        & \textbf{5.11} 
        & \textbf{0.45} 
        & \textbf{-3.69} 
        & \textbf{4.11} 
        & \textbf{4.28 $\pm $0.08} \\
        \Xhline{1.2pt}
    \end{tabular}
    }
\end{table*}

\section{Experimental Setups}
\subsection{Datasets}
We conducted all experiments using the Motus RIR dataset~\cite{gotz2021dataset} together with the VCTK clean speech corpus \cite{yamagishi2019cstr}. 
Motus is a real-world higher-order Ambisonic RIR dataset measured in a single room under 830 different furniture configurations. 
For each configuration, four fixed source--receiver position pairs were recorded, resulting in a total of 3320 raw RIR recordings. 
The measurements were acquired using an Eigenmike microphone array and stored as 32-channel RIRs, each with an effective duration of approximately 2.5~s.
Since the employed neural RIR compressor is monaural, each 32-channel RIR was decomposed into 32 single-channel RIRs, resulting in a total of 106,240 mono RIR samples. 
These samples were then randomly divided into training, validation, and test sets in an approximate ratio of 80\%/10\%/10\%.
To match the configuration of the EnCodec backbone adopted in the compressor, all RIR signals and VCTK speech signals were downsampled to 24~kHz (i.e., $s_r=24000$).

\subsection{Implementation Details} 

The model configuration of the neural RIR compressor was given in Section~\ref{sec:II-A}, where the VQ codebook size was set to 1024 (i.e., $M=1024$) and the waveform-to-token downsampling ratio is 640 (i.e., $R=640$), resulting in a bitrate of 375~bps. 
Regarding the RIR-level constraints, five short-time windows were used for computing the local energy loss (i.e., $J=5$). 
Regarding the reverberant-speech-level constraints, six configurations were used to compute the mel-domain loss (i.e., $I=6$), all of which produced 64-bin mel-spectrograms. 
At the $i$-th scale ($i=1,2,\dots,6$), the window size, hop length, and FFT size were set to $2^{i+5}$, $2^{i+5}/4$, and $\max(2^{i+5},512)$, respectively. 
The loss hyperparameters were set as $\lambda_{\mathrm{EDC}}=1$, $\lambda_{\mathrm{eng}}=10$, $\lambda_{\mathrm{adv}}=1$, $\lambda_{\mathrm{fm}}=1$, $\lambda_{q}=1000$, and
$\lambda_{\mathrm{wav}}=100$. 
Model Training was performed using the AdamW optimizer with $\beta_1=0.5$ and $\beta_2=0.9$, a batch size of 10, and 240k training iterations. 
The learning rate was initialized to 0.0003 and decayed by a factor of 0.999 after each epoch. 

    
    

\subsection{Baselines and Evaluation Metrics}

We compared the proposed method with two EnCodec-based \cite{defossez2022high} baselines: \textbf{Speech-trained EnCodec}, which was trained on the VCTK dataset using the original EnCodec training objective, and \textbf{RIR-trained EnCodec}, which is trained on the Motus RIR dataset using the same original EnCodec training objective. 
The baselines used the same architectural configuration as our neural RIR compressor to ensure a fair comparison at the same bitrate.

To comprehensively evaluate both the reconstructed RIR itself and its downstream acoustic effect, we report evaluation metrics at both the RIR level and the reverberant-speech level. 
At the RIR level, we adopted several commonly used reverberation-related evaluation metrics , including T60 error \cite{ratnarajah2024avrir}, spectral flatness measure (SFM)  error \cite{jungmann2014joint}, and direct-to-reverberant ratio (DRR) error \cite{ratnarajah2024avrir}. 
At the reverberant-speech level, the decoded and reference RIRs were separately convolved with clean speech, and the quality of the resulting reverberant speech was then evaluated.
The objective reverberant-speech evaluation metrics included log-spectral distortion (LSD), segmental signal-to-noise ratio (SegSNR), and ViSQOL \cite{chinen2020visqol}. 
We adopted mean opinion score (MOS) for the subjective evaluation of reverberant-speech naturalness.
For each MOS test, 20 test utterances generated by each compared method were included, and the evaluation was conducted by at least 25 native English listeners recruited via Amazon Mechanical Turk. 
Listeners were asked to rate the naturalness of each utterance on a 1--5 scale with a 0.5-point interval.

\section{Experimental Results}

\subsection{Main Results}

\begin{figure}[t]
    \centering
    \includegraphics[width=\linewidth]{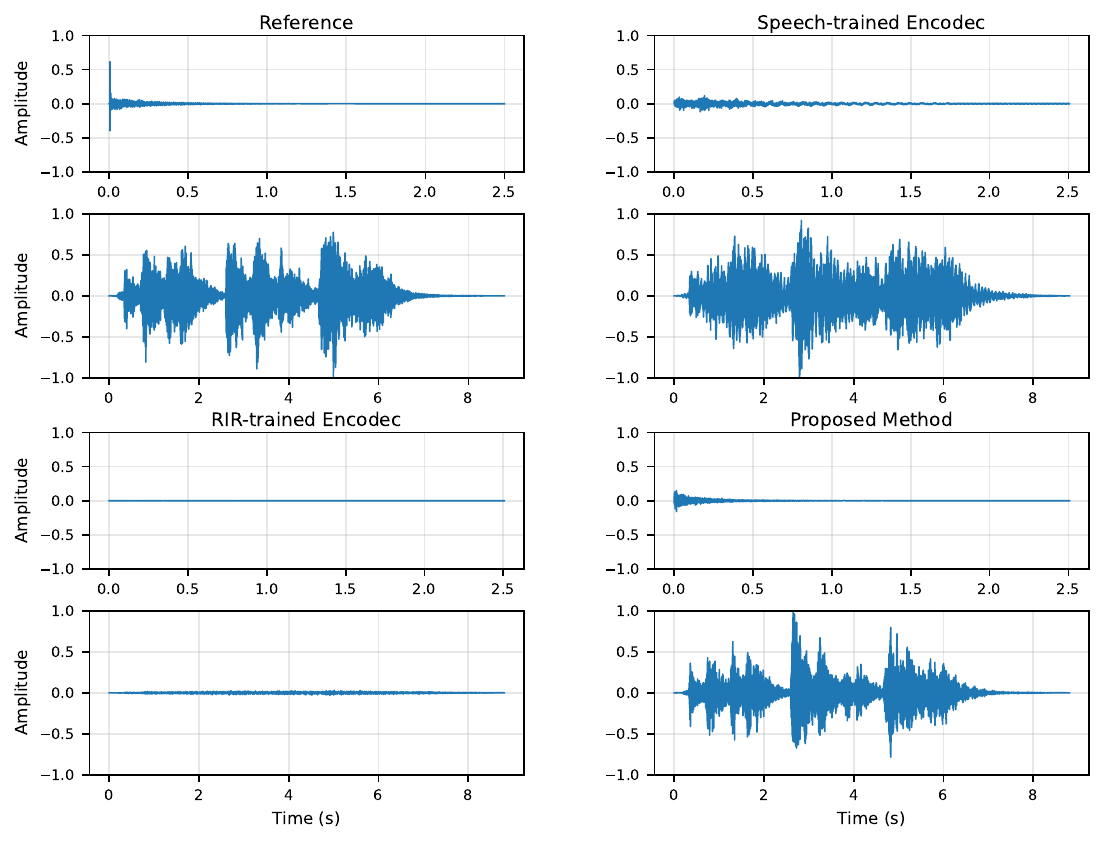}
    \caption{Visualization of RIRs (upper panel in each subfigure) and their corresponding reverberant speech waveforms (lower panel in each subfigure).}
    \label{fig:rev_speech_compare}
\end{figure}

The objective and subjective experimental results comparing the proposed method with the two baselines are summarized in Table~\ref{tab:objective_results}. 
We also plot the RIR waveforms of all methods and the reference, together with their corresponding reverberant-speech waveforms, in Fig.~\ref{fig:rev_speech_compare}. 
In the following, we analyzed the results from two perspectives. 


\textbf{RIR-Level Analysis.} 
As shown in Table~\ref{tab:objective_results}, the proposed method consistently achieved the best performance on all three RIR-level metrics. 
Although Speech-trained EnCodec was trained on speech data, it still retained some ability to roughly reconstruct the RIR, as evidenced in Fig.~\ref{fig:rev_speech_compare}. 
In contrast, RIR-trained EnCodec, despite being trained on RIR data, remained unsatisfactory on all metrics due to the lack of constraints tailored to the structural characteristics of RIRs, and failed to recover even the basic waveform shape of the RIR. 
By comparison, the proposed method yielded the lowest errors and reconstructed RIRs whose shapes more closely matched the reference signals, further demonstrating the effectiveness of the proposed structure-aware constraints in preserving the decay behavior and relative energy distribution of RIRs.

\begin{table*}[t]
\caption{Ablation study results of the proposed method.}
\label{tab:ablation}
\centering
\small
\renewcommand{\arraystretch}{1.12}
\setlength{\tabcolsep}{4.5pt}
\begin{tabular}{lccc|ccc}
\hline
Method & T60 Error (s)$\downarrow$ & SFM Error$\downarrow$ & DRR Error (dB)$\downarrow$ & LSD$\downarrow$ & SegSNR (dB)$\uparrow$ & ViSQOL$\uparrow$ \\
\hline
Proposed                  & 1.14 & 0.06 & 5.11 & 0.45 & -3.69 & 4.11 \\
\quad w/o $\mathcal{L}_{\mathrm{EDC}}$  & 4.74 & 0.07 & 7.04 & 0.46 & -4.13 & 4.06 \\
\quad  w/o $\mathcal{L}_{\mathrm{eng}}$  & 3.47 & 0.07 & 6.39 & 0.46 & -3.79 & 4.07 \\
\quad  w/o $\mathcal{L}_{\mathrm{rev-spe}}$  & 1.24 & 0.08 & 5.99 & 0.45 & -3.44 & 4.10 \\
\hline
\end{tabular}
\end{table*}

\begin{table*}[t]
\caption{Experimental results of the proposed method at different bitrates.}
\label{tab:low_bitrate}
\centering
\small
\renewcommand{\arraystretch}{1.12}
\setlength{\tabcolsep}{4.5pt}
\begin{tabular}{lccc|ccc}
\hline
Bitrate & T60 Error (s)$\downarrow$ & SFM Error$\downarrow$ & DRR Error (dB)$\downarrow$ & LSD$\downarrow$ & SegSNR (dB)$\uparrow$ & ViSQOL$\uparrow$ \\
\hline
375 bps    & 1.14 & 0.06 & 5.11 & 0.45 & -3.69 & 4.11 \\
187.5 bps  & 2.13 & 0.06 & 7.28 & 0.44 & -3.49 & 4.14 \\
93.75 bps  & 1.53 & 0.07 & 5.44 & 0.44 & -3.59 & 4.15 \\
46.875 bps & 8.89 & 0.09 & 4.74 & 0.49 & -4.72 & 3.98 \\
\hline
\end{tabular}
\end{table*}

\textbf{Reverberant-Speech-Level Analysis.} 
As shown in Table~\ref{tab:objective_results}, in line with the findings at the RIR level, the proposed method consistently outperformed the two baselines on both the objective and subjective metrics at the reverberant-speech level. 
At an ultra-low bitrate of only 375~bps, the reverberant speech generated using the RIRs decoded by the proposed method achieved a high ViSQOL score of 4.11 (with a full score of 5). 
Fig.~\ref{fig:rev_speech_compare} also shows that the reverberant speech generated by the proposed method more closely matched the reference signal in the time domain. 
This further demonstrates that the proposed structure-aware constraints enabled high-fidelity compression and reconstruction of RIRs, thereby improving the quality of the generated reverberant speech.


\subsection{Ablation Studies}

To evaluate the contributions of key losses in our proposed method, we conducted several ablation experiments. 
The results are reported in Table~\ref{tab:ablation}. 
We first ablated the EDC loss (i.e., w/o $\mathcal{L}_{\mathrm{EDC}}$), which resulted in a substantial increase in the T60 error, indicating that explicit EDC-curve supervision is crucial for preserving the global reverberation decay behavior of RIRs. 
Next, we ablated the local energy loss (i.e., w/o $\mathcal{L}_{\mathrm{eng}}$). 
The resulting increase in both T60 and DRR errors indicates that the local energy constraint plays an important role in preserving the temporal energy distribution of RIRs, particularly the relative balance among the direct sound, early reflections, and late reverberation. 
Finally, we removed all reverberant-speech-level constraints (i.e., w/o $\mathcal{L}_{\mathrm{rev-spe}}$). 
Interestingly, this led to slight degradation in several RIR-level metrics, such as the T60 error and DRR error, while causing no significant deterioration in reverberant-speech quality. 
This suggests that the reverberant-speech-level constraints can indirectly promote high-fidelity RIR reconstruction, although their contribution is less substantial than that of the losses imposed directly on RIR.



\subsection{Lower-Bitrate Limit Analysis}

As a representation of an acoustic environment, an RIR signal is primarily governed by physical factors such as propagation paths, material properties, and room modes, rather than semantic content. 
Compared with general audio signals, it is subject to stronger physical constraints and a more restricted underlying structure. 
This suggests that RIRs may admit effective compression at lower bitrates than general audio signals.
To investigate the practical lower bitrate limit for RIR compression, we further reduced the bitrate beyond 375~bps to examine the performance of decoded RIRs and reconstructed reverberant speech. 
Specifically, the bitrate was reduced from 375~bps to 187.5~bps, 93.75~bps, and 46.875~bps by changing the strides of the downsampling layers from $(4,4,5,8)$ to $(4,4,5,16)$, $(4,4,10,16)$, and $(4,8,10,16)$, respectively.

The experimental results are listed in Table~\ref{tab:low_bitrate}. 
This indicates that when the bitrate was reduced to 93.75~bps, although the RIR-level metrics degraded to some extent, the quality of the reconstructed reverberant speech could still be largely maintained. 
When the bitrate was further reduced to 46.875~bps, the metrics at both the RIR level and the reverberant-speech level deteriorated markedly, suggesting that the discrete representation had reached its capacity limit.
Therefore, the practical lower bitrate limit for RIR compression was likely to lie between 46.875~bps and 93.75~bps.
Although RIRs had a simpler structure than general audio signals and therefore offered greater potential for low-bitrate compression, effective RIR compression still could not directly rely on an audio-oriented codec framework, but instead required constraints tailored to the characteristics of RIRs.

\section{Conclusion}

\label{sec:majhead}
This paper presents a neural RIR compression method within the EnCodec framework, with structure-aware constraints jointly introduced at both the RIR level and the reverberant-speech level.
Specifically, the proposed method explicitly constrains the EDC and local energy distribution of decoded RIRs, while further promoting the consistency of the downstream reverberant speech generated from the RIRs.
These constraints help the model better capture the decay characteristics and temporal structure of RIRs, while better preserving their acoustic rendering effect.
Experimental results show that, at low bitrates, the proposed method preserves key acoustic properties of RIRs more accurately than audio-oriented codecs.
Extending this framework to spatial and multi-channel RIR scenarios will be part of our future work.











\printbibliography

\end{document}